\documentclass[prl,aps,twocolumn,,showpacs,floatfix]{revtex4}

\usepackage{amsmath,amssymb,mathrsfs}
\usepackage{graphicx}
\usepackage{version}
\DeclareGraphicsExtensions{.eps,.ps}

\begin{document}

\title
{Collective Excitations of a Two-Component Bose Condensate
at Finite Temperature}
\author{C.-H. Zhang}
\affiliation{Department of Physics, Indiana University, 727 E.\ 3rd
Street, Bloomington, IN 47405}
\author{H. A. Fertig}
\affiliation{Department of Physics, Indiana University, 727 E.\ 3rd
Street, Bloomington, IN 47405}
\date{\today}

\begin{abstract}

We compare the collective modes for Bose-condensed systems with two
degenerate components with and without spontaneous intercomponent
coherence at finite temperature using the time-dependent Hartree-Fock
approximation. We show that the interaction between the condensate
and non-condensate in these two cases results in qualitatively different
collective excitation spectra.   We show that at zero temperature
the single-particle excitations of the incoherent Bose condensate can be probed
by intercomponent excitations.

\end{abstract}

%%%%%%%%%%%%%%%%%%%%%%%%%%%%%%%%%%%%%%%%%%%%%%%%%%%%%%%%%%%%%%%%%%%%%%%%%%%%%%%
\pacs{
03.75.Fi,       % Bose-Einstein condensation
03.75.Kk,        % Dynamic properties of condensates; collective and hydrodynamic
                 % excitations, superfluid flow
67.40.Db         % Quantum statistical theory; ground state, elementary excitations
 }
%%%%%%%%%%%%%%%%%%%%%%%%%%%%%%%%%%%%%%%%%%%%%%%%%%%%%%%%%%%%%%%%%%%%%%%%%%%%%%%

\maketitle

{\em Introduction}. For a Bose-condensed system with two internal
components, one of the fundamental questions is whether one can
distinguish a system with spontaneous intercomponent
coherence from one without it\ \cite{comment}. At zero temperature, the
ground state energies calculated with the static Gross-Pitaevskii
energy functional are the same for both cases\ \cite{Pethick02}. The
collective excitations calculated with the linearized time-dependent
Gross-Pitaevskii (TDGP) equation are also identical for both cases.
For example, the collective modes for a homogeneous two-component
system are
\begin{align} \label{eq:dispers_T0}
\omega^2_{\pm}=\frac{\omega_1^2+\omega_2^2
\pm\sqrt{\left(\omega_1^2-\omega_2^2\right)^2
+16g^2_{12}\rho_1\rho_2(\varepsilon^0_{\vec{q}})^2}}{2}
\end{align}
for both cases, as obtained from the linearized TDGP\ \cite{Timmermans98}
or the Bogoliubov theory \ \cite{Goldstein97,Search01,Tommasini03},
where $\varepsilon^0_{\vec{q}}=\frac{\hbar^2\vec{q}^2}{2m}$,
$\omega^2_i=\varepsilon^0_{\vec{q}}\left(\varepsilon^0_{\vec{q}}
+2g_{ii}\rho_i\right)$ is the Bogoliubov mode for a single component
system, $g_{ii}$ is the $i$-th intracomponent contact interaction,
$g_{12}$ the intercomponent contact interaction, and $\rho_i$ is the
density of $i$-th component with $i=1,2$. The
excitations in Eq. (\ref{eq:dispers_T0}) may be interpreted
differently for the incoherent and
coherent cases are different. 
Without intercomponent coherence, there are {\em two} macroscopically
occupied states, a situation that has come to be known as a fragmented
condensate \cite{mueller06}.  In this case
$\omega_{+}$ may be interpreted as a density wave in which the two condensates
move in phase, while $\omega_{-}$ is a collective excitation
associated with out of phase motion between the two condensates.
With intercomponent coherence, there is only a {\em single}
condensate. In this case $\omega_{+}$ is interpreted
as the superfluid mode, and $\omega_-$ is
a Goldstone mode corresponding to a spontaneously broken U(1) symmetry
in the relative phase between the two components. As
pointed out by Leggett\ \cite{comment}, in the absence of particle
exchange between the two components, no physical quantities
can depend on the relative phase between the them.  
However, at finite temperatures, we will show that the coupling
between the condensate and non-condensate particles provides a way of
probing the nature of the condensates. 

\begin{figure}[t]
\includegraphics[width=1\columnwidth]{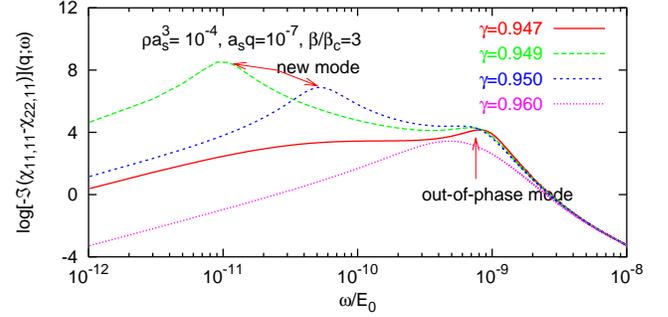}
\vspace{-5mm} \caption{(Color online) The appearing of the new
out-of-phase mode in the incoherent two-component systems at finite
temperature.} \label{fig:im_chist3}
\end{figure}

\begin{figure}[t]
\includegraphics[width=1\columnwidth]{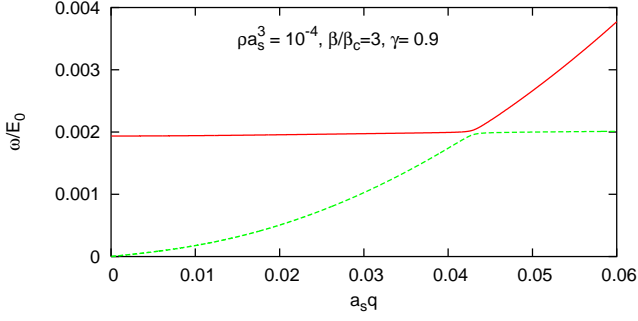}
\vspace{-5mm} \caption{(Color online) The two pseudo-spin modes in
the system with intercomponent coherence.} \label{fig:disper}
\end{figure}

In this work, we compare the collective excitations for a
two-component Bose-condensed system with and without intercomponent
coherence at finite temperature using the time-dependent
Hartree-Fock (TDHF) approximation, for the special case
$g_{11}=g_{22}>g_{12}$. Our principal findings are as
follows. Without intercomponent coherence: (i) The motions of the
two components and uncondensed particles give rise to gapless
collective excitations. We find one in-phase (superfluid) mode. However, there
can be {\it two} out-of-phase modes within some range of 
intercomponent interaction strengths at appropriate temperatures.
This is illustrated in Fig.\ \ref{fig:im_chist3}. (ii) 
The perturbative response to a field that allows intercomponent
conversion has a resonance at frequencies
given by the
single-particle excitation energies. With
intercomponent coherence, we find a gapless superfluid mode, and
two ``pseudo-spin waves'', one gapless and the other gapped,
as illustrated in Fig.\ \ref{fig:disper}. The gapped pseudo-spin wave
%arises because of spin flipping between two non-condensate
%spinor states, and
has
vanishing weight as temperature goes to zero.  The differences in
in these spectra allow one to distinquish a single condensate from 
a fragmented one.

{\em Model}. We consider the collective excitations in a
two-component homogenous Bose-condensed atomic gas at finite
temperatures $k_BT=1/\beta$. To simplify the physical picture, we
assume the two internal states are degenerate and take the two-body
interactions as contact forms with strength parameters
$g_{11}=g_{22}=g_{s}=\frac{4\pi\hbar^2}{m}a_s$ and
$g_{12}=g_x=\frac{4\pi\hbar^2}{m}a_x$ with $a_{s,x}$ the
corresponding $s$-wave scattering length, and take $g_s>g_x$, so
that the homogenous state is stable against phase separation.
The system may be characterized by the gas parameter
$\rho a^3_s$, where $\rho$ is the total particle density, and
$\gamma=g_x/g_s$.  Our unit of energy is $E_0=\frac{\hbar^2}{2m a_s^2}$,
%and all
%quantities can be expressed in combinations of $\rho a^3_s$,
%$E_0=\frac{\hbar^2}{2m a_s^2}$, and $\gamma=g_x/g_s$, with $\hbar$
%the Planck constant and 
where $m$ the mass of the particles. Using a
pseudo-spin language, the single-particle wavefunctions 
may be written as spinors with components
$\psi_{\vec{k}\lambda}(\vec{r})=V^{-\frac12}e^{i\vec{k}\cdot\vec{r}}\eta_{\lambda}$,
with $V$ the volume of the system and $\eta_{\lambda}$ the spin
state functions. In this work, the spinor index is
denoted by Greek letters.  For incoherent condensates, we write
$\lambda=1,2$ with $\eta_1=\left(\begin{array}{c}
1\\0\end{array}\right)$ and $\eta_2=\left(\begin{array}{c}
0\\1\end{array}\right)$ the pseudospin states of the two condensates.
For coherent case, we write $\lambda=\pm$, 
$\eta_+=\frac{1}{\sqrt{2}}\left[\eta_1+ \eta_2\right]$, 
$\eta_-=\frac{1}{\sqrt{2}}\left[\eta_1- \eta_2\right]$,
and the condensation is assumed to be in the $\eta_+$ state.
In both cases the macroscopically occupied states have zero momentum.

{\em Method}. We briefly outline our method here;
details will be presented elsewhere. It is well known that to
treat the dynamics of a condensate and thermally excited (non-condensate)
particles in a fully consistent manner is theoretically challenging\
\cite{Griffin96}. Recently we have shown this can be done within a TDHF
approximation\ \cite{Zhang06} by using a constrained grand canonical
ensemble\ \cite{Huse82} in which the condensate particle
number is fixed.  We first determine the static
Hartree-Fock (HF) ground state properties, including the chemical
potential $\mu$, the single-particle energies
$\varepsilon_{\vec{k}\alpha}$, the occupation numbers
$n_{\vec{k}\alpha}$, and the critical temperature $\beta_c$.
This involves defining a self energy $\Sigma$ which is a functional
of the densities $\rho_{\vec{k}_1\lambda_1,\vec{k}_2\lambda_2} \equiv
<a^{\dag}_{\vec{k}_1\lambda_1} a_{\vec{k}_2\lambda_2}>$,
with $a_{\vec{k}\lambda}$ the boson destruction operator,
and the condensate wavefunction(s) $\psi_{0\lambda}$.
We then introduce a small perturbation $\delta U$, and assume
$\Sigma$ keeps its functional form with respect to the $\rho$'s and $\psi_{0\lambda}$,
which are now time-dependent.  This results in self-consistent 
equations which can be solved in the linear response regime.
This allows us to obtain equations for the response functions
$\chi^{\vec{k}_1\vec{k}_2}_{\lambda\mu,\alpha\beta}(\vec{q};\omega)
=i\int_0^{\infty}\frac{e^{i\omega t}dt}{2\pi}\langle
a^\dagger_{\vec{k}_1+\vec{q}\mu}(t)a_{\vec{k}_1\lambda}(t)
a^\dagger_{\vec{k}_2\alpha}(0)a_{\vec{k}_2+\vec{q}\beta}(0)\rangle$,
\begin{widetext}
\begin{align}
\label{eq:chi} \left(\omega+\varepsilon_{\vec{k}+\vec{q}\mu}
-\varepsilon_{\vec{k}\lambda}\right)\chi^{\vec{k}
\vec{k}^\prime}_{\lambda\mu,\alpha\beta} (\vec{q};\omega)
&=(n_{\vec{k}+\vec{q}\mu}-n_{\vec{k}\lambda})
\delta_{\alpha\lambda}\delta_{\beta\mu}\delta_{\vec{k}\vec{k}^\prime}
+(n_{\vec{k}+\vec{q}\mu}-n_{\vec{k}\lambda})
\sum_{ss^\prime,\lambda_1\mu_1,\vec{k}_1}K^{ss^\prime;\vec{k}_1}_{\lambda\mu_1\lambda_1\mu}
\chi^{\vec{k}_1\vec{k}^\prime}_{\lambda_1\mu_1,\alpha\beta}(\vec{q};\omega)
\end{align}
with a vertex
\begin{align}
\label{eq:vertex}
K^{ss^\prime;\vec{k}}_{\lambda\mu_1\lambda_1\mu}=g^{ss^\prime}_{\lambda\mu_1\lambda_1\mu}
\left(1-\delta_{\lambda c}\delta_{\lambda_1c}\delta_{\vec{k},0}
-\delta_{\mu c}\delta_{\mu_1c}\delta_{\vec{k},-\vec{q}}\right)
+g^{ss^\prime}_{\lambda\mu_1\mu\lambda_1}\left[1-\left(\delta_{\lambda
c}+\delta_{\mu c}\right)\left(\delta_{\lambda_1c}\delta_{\vec{k},0}
+\delta_{\mu_1c}\delta_{\vec{k},-\vec{q}}\right)\right].
\end{align}
\end{widetext}
Here $s$ and $s^\prime$ denote the components of each spinor, and
$g^{ss^\prime}_{\lambda\mu_1\lambda_1\mu}=\frac{g_{ss^\prime}}{V}\eta_{\lambda}(s)
\eta_{\mu_1}(s^\prime)\eta_{\lambda_1}(s^\prime)\eta_{\mu}(s)$. The
Kronecker delta terms $\delta_{\lambda c}$ are 1 when $\lambda$ is a condensate
state, and 0 otherwise.  Poles of the response functions correspond
to collective excitations of the system.

{\em Results.} We first consider incoherent condensates. In this case, the
density-density response function matrix has only six non-zero
matrix elements:
$\chi_{11,11}(\vec{q};\omega)=\chi_{22,22}(\vec{q};\omega)$,
$\chi_{11,22}(\vec{q};\omega)= \chi_{22,11}(\vec{q};\omega)$,
and
$\chi_{12,21}(\vec{q};\omega)=\chi_{21,12}(-\vec{q};-\omega)$, where
$\chi_{\lambda\mu,\alpha\beta}(\vec{q};\omega)=\frac{1}{V}
\sum_{\vec{k}_1,\vec{k}_2}\chi^{\vec{k}_1\vec{k}_2}
_{\lambda\mu,\alpha\beta}(\vec{q};\omega)$. The equations for the
first four functions are decoupled from those of the latter two.

\begin{figure}[t]
\includegraphics[width=1\columnwidth]{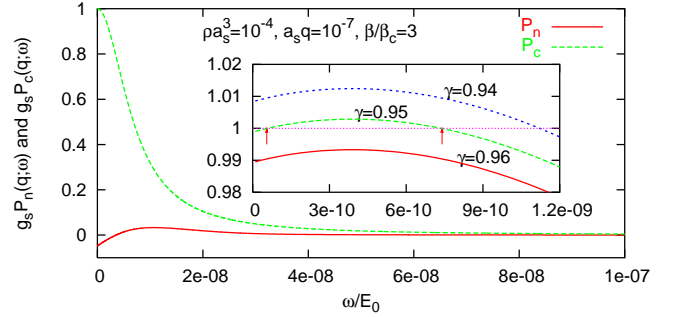}
\vspace{-5mm} \caption{(Color online) The condensate and
non-condensate polarization functions at which the new out-of-phase
mode appears. The inset shows $g_s(2-\gamma)P(\vec{q};\omega)$. The
arrows show the locations of the two out-of-phase modes.}
\label{fig:polar}
\end{figure}

The functions $\chi_{11,11}(\vec{q};\omega)$ and
$\chi_{22,11}(\vec{q};\omega)$ describe the density response of the two
components. From Eqs.\ (\ref{eq:chi}) and (\ref{eq:vertex}),
one obtains
\begin{align}
\label{eq:chi_incoh} \left(\begin{array}{cc} 1-2g_{s}P
&-g_{x}P\\-g_{x}P& 1-2g_{s}P
\end{array}\right)\left(\begin{array}{c}
\chi_{11,11}\\\chi_{22,11}
\end{array}\right)
=\left(\begin{array}{c} P\\0
\end{array}\right),
\end{align}
where $P(\vec{q};\omega)=P_c(\vec{q};\omega)+P_n(\vec{q};\omega)$
with $P_c(\vec{q};\omega)=\frac{2\rho_0\varepsilon^0_{\vec{q}}}
{\omega^2-(\varepsilon^0_{\vec{q}})^2
+2g_{s}\rho_0\varepsilon^0_{\vec{q}}}$ and $P_{n}(\vec{q};\omega)
=\frac{1}{\Omega}\sum_{\vec{k}\ne0,-\vec{q}}\frac{n_{\vec{q}+\vec{k}}-n_{\vec{k}}}
{\omega+\varepsilon^0_{\vec{k}+\vec{q}}-\varepsilon^0_{\vec{k}}}$,
with $n_{\vec{k}}=n_{\vec{k}1}=n_{\vec{k}2}$. The poles of these $\chi$'s
occur when the determinant of the matrix in Eq. (\ref{eq:chi_incoh}) vanishes,
\begin{align}
\label{eq:pole-incoh1} 1-(2g_s\pm
g_x)\left[P_c(\vec{q};\omega)+P_{n}(\vec{q};\omega)\right]=0.
\end{align}
Equation (\ref{eq:pole-incoh1}) shows the
effect of intercomponent interaction and the interaction between the
condensate and non-condensate on the collective excitations. The
plus and minus signs in Eq.\ (\ref{eq:pole-incoh1}) can be
interpreted as resulting from the in-phase and out-of-phase motion
of the two components, respectively. We find only one in-phase
(gapless) mode as a solution, which reduces to $\omega_+$ in Eq.\
(\ref{eq:dispers_T0}) in the zero temperature limit. 
Thus the interaction between the
condensate and non-condensate has little effect on the in-phase
motion. However, this interaction has a striking effect on  the
out-of-phase motion. Within a range $g_{x,1}<g_x<g_{x,2}$, the
values $g_{x,1}$ and $g_{x,2}$ depending on the gas parameter $\rho
a^3$ and temperature $T$, the system can support a new out-of-phase
mode for small $\vec{q}$.
This is illustrated
Fig.\
\ref{fig:im_chist3} for
$\rho a^3=10^{-4}$ and $\beta/\beta_c=3$,
where it can be seen that
$\frac12\Im\left[\chi_{1111}(\vec{q};\omega)
-\chi_{2211}(\vec{q};\omega)\right]$ carries an extra resonance.
In order to see why this occurs, we show the
polarization functions $P_c$ and $P_n$ in Fig.\ \ref{fig:polar}. For
$\varepsilon^0_{\vec{q}}<2g_s\rho_0$,
$P_c(\vec{q};\omega)$ is a monotonic function of $\omega$
and approaches $1/g_s$ with zero slope as $\omega\rightarrow0$,
while $P_n(\vec{q};\omega)$ is non-monotonic at small $\omega$. 
The combination $g_s(2-\gamma)P(\vec{q};\omega)$ is then
non-monotonic at small $\omega$, 
so for certain ranges of $\gamma$, Eq. \ref{eq:pole-incoh1}
is satisfied twice.  This is shown
in the inset in Fig.\
\ref{fig:polar}.

The function $\chi_{21,12}(\vec{q};\omega)$ supports a 
collective mode\ \cite{Oktel99,Johnsen01,Pethick01}, corresponding
to an excitation in which the ``flavor'' of a
particle changes.
From
Eq.\;(\ref{eq:chi}) one finds
\begin{align}
\label{eq:chi_2112}
\chi_{21,12}(\vec{q};\omega)%=\chi_{1221}(-\vec{q};-\omega)
=\frac{P^c_{12}(\vec{q};\omega)
+P^{n}(\vec{q};\omega)}{1-g_x\left[P^c_{12}(\vec{q};\omega)
+P_{n}(\vec{q};\omega)\right]},
\end{align}
where
\begin{align}
P^c_{12}(\vec{q};\omega)=\frac{2\rho_0\left(\varepsilon^0_{\vec{q}}+
g_s\rho_0\right)}{\omega^2-\left(\varepsilon^0_{\vec{q}}+
g_s\rho_0\right)^2+2g_x\rho_0\left(\varepsilon^0_{\vec{q}}+
g_s\rho_0\right)},
\end{align}
At zero temperature the collective mode 
requires exciting a particle from one of the condensates
to an excited state of the other component, so
$\chi_{21,12}$ has a pole
at the single particle excitation spectrum,
$\omega=\varepsilon_{\vec{q}}-\mu=\varepsilon^0_{\vec{q}}+g_s\rho_0$.  
Thus, this
provides a way to probe the single-particle spectrum. 
We note that, since $\chi_{21,12}$ is the response to
a ``flavor-changing'' perturbation, this collective mode is
{\it not} found in the TDGP approximation when species exchange
is absent from the unperturbed Hamiltonian.  It is apparent that
our TDHF approach is able to capture a broader variety of excitations
for this system.

We now turn to the coherent case.  The system can condense into any
spinor given by
$\eta(\theta)=\frac{1}{\sqrt{2}}\left(\eta_1+e^{i\theta}\eta_2\right)$,
resulting in a single condensate. We choose the system condensing into
$\eta_+=\eta(\theta=0)$, which spontaneously breaks the U(1)
symmetry. Therefore, we would expect at least two gapless modes: a
superfluid density mode and a Goldstone mode (a pseudo-spin wave)
for the spontaneous U(1) symmetry breaking. 

The density wave may be found from the poles of
$\chi_{++}(\vec{q};\omega)=\sum_{\vec{k}_1\vec{k}_2;\alpha\beta}
\chi^{\vec{k}_1\vec{k}_2}_{++,\alpha\beta}(\vec{q};\omega)$ and
$\chi_{--}(\vec{q};\omega)=\sum_{\vec{k}_1\vec{k}_2;\alpha\beta}
\chi^{\vec{k}_1\vec{k}_2}_{--,\alpha\beta}$, which follow a matrix
equation obtained from Eq.\ (\ref{eq:chi}),
\begin{align}
\left(\begin{array}{cc}
1-g_{p}P_{++}&-g_{s}P_{++}\\
-g_{s}P_{--} &1-g_{p}P_{--}
\end{array}\right)\left(\begin{array}{c}
\chi_{++}\\\chi_{--}\end{array}\right) =\left(\begin{array}{c}
P_{++}\\P_{--}\end{array}\right),
\end{align}
where $g_{p}=g_s+g_x$,
$P_{++}(\vec{q};\omega)=P_c^{++}(\vec{q};\omega)+P_n^{++}(\vec{q};\omega)$,
$P_{--}(\vec{q};\omega)=P^n_{--}(\vec{q};\omega)$ with
$P_c^{++}(\vec{q};\omega)=\frac{2\rho_0\varepsilon_{\vec{q}}}
{\omega^2-(\varepsilon^0_{\vec{q}})^2
+g_{p}\rho_0\varepsilon^0_{\vec{q}}}$ and
$P_{n}^{\alpha\beta}(\vec{q};\omega)
=\frac{1}{\Omega}\sum_{\vec{k}}^\prime\frac{n_{\vec{q}+\vec{k}\alpha}-n_{\vec{k}\beta}}
{\omega+\varepsilon_{\vec{q}+\vec{k}\alpha}-\varepsilon_{\vec{k}\beta}}$,
where the prime excludes the momentum of the condensate modes.
As in the single component case, the 
non-condensate is too dilute to sustain a propagating second sound
mode within this approximation.  The resulting response and superfluid
mode are thus qualitatively similar to the single component case.

By contrast, the non-condensate component has a dramatic effect on the pseudo-spin
waves. In order to see this, we define
$\chi^{\vec{k}_1}_{\pm\mp}(\vec{q};\omega)=\sum_{\vec{k}_2\alpha\beta}
\chi^{\vec{k}_1\vec{k}_2}_{\pm\mp,\alpha\beta}(\vec{q};\omega)$ and
split these into condensate and non-condensate parts: 
$\chi^c_{+-}(\vec{q};\omega)=\chi^{\vec{k}=0}_{+-}(\vec{q};\omega)$,
$\chi^c_{-+}(\vec{q};\omega)=\chi^{\vec{k}=-\vec{q}}_{-+}(\vec{q};\omega)$,
and
$\tilde{\chi}_{\pm\mp}(\vec{q};\omega)=\sum_{\vec{k}}^\prime
\chi^{\vec{k}}_{\pm\mp}(\vec{q};\omega)$, where the prime on the sum excludes the
momentum of the condensate mode. In terms of these quantities,
Eq.\ (\ref{eq:chi}) yields
\begin{widetext}
\begin{align}
\label{eq:polepm} \left(\begin{array}{cccc}
\omega+\mu-\varepsilon_{-,\vec{q}} & -\frac12g_m\rho_0 &
-g_s\rho_0 & -g_m\rho_0\\
\frac12g_m\rho_0 &\omega+\varepsilon_{-,\vec{q}}-\mu
&g_m\rho_0 & g_s\rho_0\\
-g_sP^{+-}_n(\vec{q};\omega) & g_mP^{+-}_n(\vec{q};\omega)
&1-g_sP^{+-}_n(\vec{q};\omega) &g_mP^{+-}_n(\vec{q};\omega) \\
g_mP^{-+}_n(\vec{q};\omega) & -g_sP^{-+}_n(\vec{q};\omega) &
g_mP^{-+}_n(\vec{q};\omega) & 1-g_sP^{-+}_n(\vec{q};\omega)
\end{array}\right)
\left(\begin{array}{c}
\chi^c_{-+}(\vec{q};\omega)\\\chi^c_{+-}(\vec{q};\omega)\\\tilde{\chi}_{-+}(\vec{q};\omega)
\\\tilde{\chi}_{+-}(\vec{q};\omega)\end{array}\right)
= \left(\begin{array}{c}
\rho_0\\-\rho_0\\P^{+-}_n(\vec{q};\omega)\\P^{-+}_n(\vec{q};\omega)
\end{array}\right),
\end{align}
\end{widetext}
where $g_m=g_s-g_x$.  Eq. \ref{eq:polepm} yields 
two spin waves, as shown in
Fig.\ \ref{fig:disper}, one of which is gapless and originates from
the U(1) symmetry breaking of the relative phase of the two
internal states. At $T=0$, this gapless mode
is reduce to $\omega_-$ given by Eq.\ (\ref{eq:dispers_T0}). The
other spin wave has a gap as $q \rightarrow 0$ given by
\begin{align}
\Delta=\sqrt{g_x^2\left[\rho_0^2+(\rho_+-\rho_-)^2\right]
+g_xg_p\rho_0(\rho_+-\rho_-)}.
\end{align}
Here $\rho_+$ and $\rho_-$ are the
non-condensate density in spinor $\eta_+$ and $\eta_-$,
respectively.
The gapped spin wave, unlike
the gapped flavor-changing mode obtained from Eq.\ (\ref{eq:chi_2112}) in
the incoherent case, disappears as $T\rightarrow0$ (i.e., its weight
in the response functions vanishes), which can be see
directly from Eq.\ (\ref{eq:polepm}) since both $P^n_{+-}$ and
$P^n_{-+}$ are zero in this limit.

In the above calculations we have discussed the 
special case of interaction strengths
$g_{11}=g_{22}$. For $g_{11}\ne g_{22}$ but
$g_{12}^2<g_{11}g_{22}$ so that the homogeneous state is still
stable against phase separation, we find for the incoherent case
results which are qualitatively the same. 
The coherent system, by contrast, is
more complicated. In particular one finds that 
the spinors of the non-condensate modes
are neither parallel nor anti-parallel to the condensate mode
spinor. This leads to the response functions in Eq. \ref{eq:chi} all
being coupled together, so a classification
of the modes as flavor-changing and flavor-preserving
is no longer possible.  Nevertheless,
one still finds two gapless linear modes and one
gapped mode
as we found in the $g_{11}=g_{22}$ case.  Details of this
more complicated situation will be presented elsewhere.

In conclusion, we have compared the collective excitations of
two-component Bose-condensed systems with and without coherence at
finite temperature. The coupling between the condensate and
non-condensate depends on whether there is intercomponent coherence,
and the two identical dispersions given
by Eq.\ (\ref{eq:dispers_T0}) at zero temperature evolve to quite
different structures. We also have shown that, at zero temperature
the single-particle excitations in the incoherent case can be probed
by flavor-changing excitations, which can not be done in the coherent case.

This work was supported by the NSF Grant No. DMR0454699.

\bibliography{bec_two-component}
\end{document}